\documentstyle[twocolumn,pra,aps,amssymb,psfig]{revtex}

\newcommand{\mbb}[1]{\mbox{\boldmath $#1$}}

\begin{document}

\draft

\title{Entanglement generation and degradation by passive optical devices}
\author{S.~Scheel and D.-G.~Welsch}
\address{Theoretisch-Physikalisches Institut, 
Friedrich-Schiller-Universit\"at Jena, Max-Wien-Platz 1,
D-07743 Jena, Germany}

\date{\today}
\maketitle

\begin{abstract}
The influence of losses in the interferometric generation and
the transmission of continuous-variable entangled light is studied,
with special emphasis on Gaussian states. Based on the theory of
quantum-state transformation at absorbing dielectric devices,
the amount of entanglement is quantified by means of the
relative-entropy measure. Upper bounds of entanglement
and the distance to the set of separable Gaussian states are
calculated. Compared with the distance measure, the bounds
can substantially overestimate the entanglement. In particular,
they do not show the drastic decrease of entanglement with
increasing mean photon number, as does the distance measure. 
\end{abstract}


%%%%%%%%%%%%%%%%%%%%%%%%%%%%%%%%%%%%%%%%%%%%%%%%%%%%%%%%%%%%%%%%%%%%%%
\section{Introduction}

Entangled quantum states containing more than one photon on
average have been of increasing interest (see for example
\cite{Korolkova00}) for several reasons. So, maximally entangled
continuous-variable states of EPR type require infinitely large
mean numbers of photons. Another reason is that such states might be
more robust against decoherence. 
If, for example, in an experiment with Bell states one photon is lost,
e.g. absorbed, during the transmission from a sender to a receiver,
the entanglement is immediately gone (the state is projected onto a
separable state). 
States with more than one photon on average have the 
advantage that even a few photons might get lost on the way (leaving
behind some mixed state), but inseparability might still be preserved, 
that is, maximally entangled photon pairs could still be extracted by
means of purification procedures \cite{Duan00a,Duan00b,Parker00}.

The aim of the present article is to investigate entanglement
properties of bipartite quantum states that generically live in
infinite-dimensional Hilbert spaces. Typical examples
are Gaussian states such as two-mode squeezed vacuum states,
which are the states commonly used in quantum communication of
continuous-variable systems \cite{Braunstein98}. They  
are also the states we will be looking at in what follows.

In real experiments, both in quantum-state generation and
processing, e.g., transmission through generically lossy optical
systems such as fibers,
the necessarily existing interaction of the light fields with 
dissipative environments spoils the quantum-state purity, leaving
behind statistical mixtures.
Unfortunately, quantification of entanglement for mixed states
in infinite-dimensional Hilbert spaces is yet impossible
in practice. It typically involves minimizations over infinitely many
parameters, as it is the case for the entropy 
of formation as well as for the distance to the set of all separable
quantum states measured either by the relative entropy or Bures'
metric \cite{Vedral98}. It is, however, possible to derive upper
bounds on the entanglement content \cite{Hiroshima00} by using the
convexity property of the relative entropy. For Gaussian states,
however, we derive an upper bound based on the distance to the set of
separable Gaussian states which is far better than the convexity
bound. These bounds may be very useful for estimation of the
entanglement degradation in real quantum information systems.  

The article is organized as follows. In Section \ref{sec:gen}
the influence of losses in the interferometric entanglement
generation at a beam splitter is studied. 
A typical situation in quantum communication is considered in
Section \ref{sec:deg}, in which the entanglement degradation of a
two-mode squeezed vacuum (TMSV) state is transmitted through a noisy
communication channel say, two lossy optical fibers, is examined.
Some concluding remarks are given in Section \ref{conclusions}.

%%%%%%%%%%%%%%%%%%%%%%%%%%%%%%%%%%%%%%%%%%%%%%%%%%%%%%%%%%%%%%%%%%%%%%
\section{Entanglement generation by mixing squeezed vacua
at a beam splitter}
\label{sec:gen}
\subsection{Lossless beam splitters}

\begin{figure}[h]
\hspace{1cm}
\psfig{file=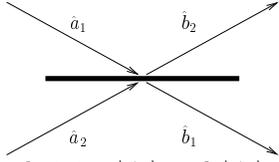,width=5cm}
\caption{\label{bs} Squeezed states \protect$|\psi_1\rangle$ and
\protect$|\psi_2\rangle$ impinging on a beam splitter producing
entangled light beams.}
\end{figure}

Let us first consider the case of a lossless beam splitter and
(quasi-)monochromatic light of (mid-)frequency $\omega$ (Fig.~\ref{bs}).
It is well known
\cite{Yurke86,Prasad87,Ou87,Fearn87,Campos89,Leonhardt93} that a
lossless beam splitter 
transforms the operators of the incoming modes
$\hat{a}_1(\omega)$ and $\hat{a}_2(\omega)$ to the operators of the
outgoing modes $\hat{b}_1(\omega)$ and $\hat{b}_2(\omega)$ according to
\begin{equation}
\label{2.1}
{ \hat{b}_1(\omega) \choose \hat{b}_2(\omega) }
= {\bf T}(\omega) 
{ \hat{a}_1(\omega) \choose \hat{a}_2(\omega) },
\end{equation}
where ${\bf T}(\omega)$ is the unitary characteristic
transformation matrix of the beam splitter.
Equivalently, the operators can be left unchanged and instead the
density operator is transformed with the inverse matrix
${\bf T}^{-1}(\omega)$ $\!=$ $\!{\bf T}^+(\omega)$
according to
\begin{equation}
\label{2.2}
\hat{\varrho}_{\rm out} = \hat{\varrho}_{\rm in}\!\left[
{\bf T}^+(\omega)
{ \hat{a}_1(\omega) \choose \hat{a}_2(\omega) }
\,,{\bf T}^{\rm T}(\omega)
{ \hat{a}_1^\dagger(\omega) \choose \hat{a}_2^\dagger(\omega) }
\right].
\end{equation}

Let each of the two incoming modes be prepared in
a squeezed vacuum state, i.e.,
\begin{equation}
\label{2.5}
\hat{\varrho}_{\rm in}=
|\Psi_{\rm in}\rangle\langle\Psi_{\rm in}|
\end{equation}
where
\begin{equation}
\label{2.3}
|\Psi_{\rm in}\rangle = \hat{S}_1\hat{S}_2 |0,0\rangle, 
\end{equation}
with $\hat{S}_i$ ($i$ $\!=$ $\!1,2$) being the (single-mode)
squeeze operator
\begin{eqnarray}
\label{2.4}
\lefteqn{
\hat{S}_i = \exp\!\left[
   -\textstyle\frac{1}{2}\left( \xi_i\hat{a}_i^{\dagger 2}
   - \xi_i^\ast \hat{a}_i^2 \right) \right]
}
\nonumber\\&&\hspace{2ex}
   = \exp\!\left(-\textstyle\frac{1}{2}q_i\hat{a}_i^{\dagger 2}\right)
   \left(1\!-\!|q_i|^2\right)^{(2\hat{n}_i+1)/4}
   \exp\!\left(\textstyle\frac{1}{2}q_i^\ast\hat{a}_i^2\right)       
\end{eqnarray}
($q_i$ $\!=$ $\!\tanh|\xi_i|e^{i\phi_i}$, $\phi_i$ $\!=$ $\!\arg \xi_i$).
Here, the second line follows from general disentangling theorems
\cite{Wodkiewicz85,Ma90}. By Eq.~(\ref{2.2}), the output quantum
state is 
\begin{equation}
\label{2.6}
\hat{\varrho}_{\rm out}= |\Psi_{\rm out}\rangle\langle\Psi_{\rm out}|,
\end{equation}
where
\begin{eqnarray}
\label{2.6a}
\lefteqn{
|\Psi_{\rm out}\rangle
   = \left[(1-|q_1|^2)(1-|q_2|^2)\right]^{1/4}
}
\nonumber\\&&\hspace{4ex}\times\,   
   \exp\!\left[ -\textstyle\frac{1}{2}q_1\big( T_{11}\hat{a}_1^\dagger
   + T_{21}\hat{a}_2^\dagger \big)^2
   \right.
\nonumber\\&&\hspace{6ex}
   \left. 
   -\,\textstyle\frac{1}{2}q_2 \big( T_{12}\hat{a}_1^\dagger
   + T_{22}\hat{a}_2^\dagger \big)^2 \right]
   |0,0\rangle.
\end{eqnarray}
The $T_{ij}$ are the elements of the characteristic
transformation matrix ${\bf T}$ (at chosen mid-frequency), which can
be given, without loss of generality, in the form of
\begin{equation}
\label{2.8}
{\bf T}
   = \left(
   \begin{array}{cc} T & R \\ -R^\ast & T^\ast
   \end{array}
   \right),
\end{equation}
with $T$ $\!=$ $\!|T|{\rm e}^{i\phi_T}$ and $R$ $\!=$
$\!|R|{\rm e}^{i\phi_R}$ being the (complex) transmission and reflection
coefficients of the beam splitter. 

{F}rom inspection of Eq.~(\ref{2.6}) it is seen that the
preparation of an entangled state is controlled by the parameter
\begin{equation}
\label{2.7}
\xi_{12} = q_1 T_{11}T_{21}+ q_2 T_{12}T_{22}
= - q_1 T R^\ast + q_2 R T^\ast .
\end{equation}
When $\xi_{12}$ $\!=$ $\!0$ is valid, then the output state is
separable. This is the case
for \mbox{$\phi_1$ $\!-$ $\!\phi_2$ $\!+$ $\!2(\phi_T$
$\!-$ $\!\phi_R)$ $\!=$ $\!0$} and \mbox{$|q_1|$ $\!=$ $\!|q_2|$}.
On the other hand, if again \mbox{$|q_1|$ $\!=$ $\!|q_2|$
$\!=$ $\!|q|$} but \mbox{$\phi_1$ $\!-$ $\!\phi_2$ $\!+2(\phi_T$
$\!-\phi_R)$ $\!=$ $\!\pm\pi$}, then for $|TR|$ $\!=$ $\!1/2$
the output quantum state is just a TMSV state,
\begin{equation}
\label{2.8a}
|\Psi_{\rm out}\rangle
= |{\rm TMSV}\rangle =\sqrt{1-|q|^2}\,
\exp\!\left[-q\hat{a}_1^\dagger\hat{a}_2^\dagger \right]
|0,0\rangle,
\end{equation}
where
\begin{equation}
\label{2.9}
q = |q|^{i(\phi_2+\phi_R-\phi_T)}
= -|q| {\rm e}^{i(\phi_1+\phi_T-\phi_R)}. 
\end{equation}

\begin{figure}[h]
\psfig{file=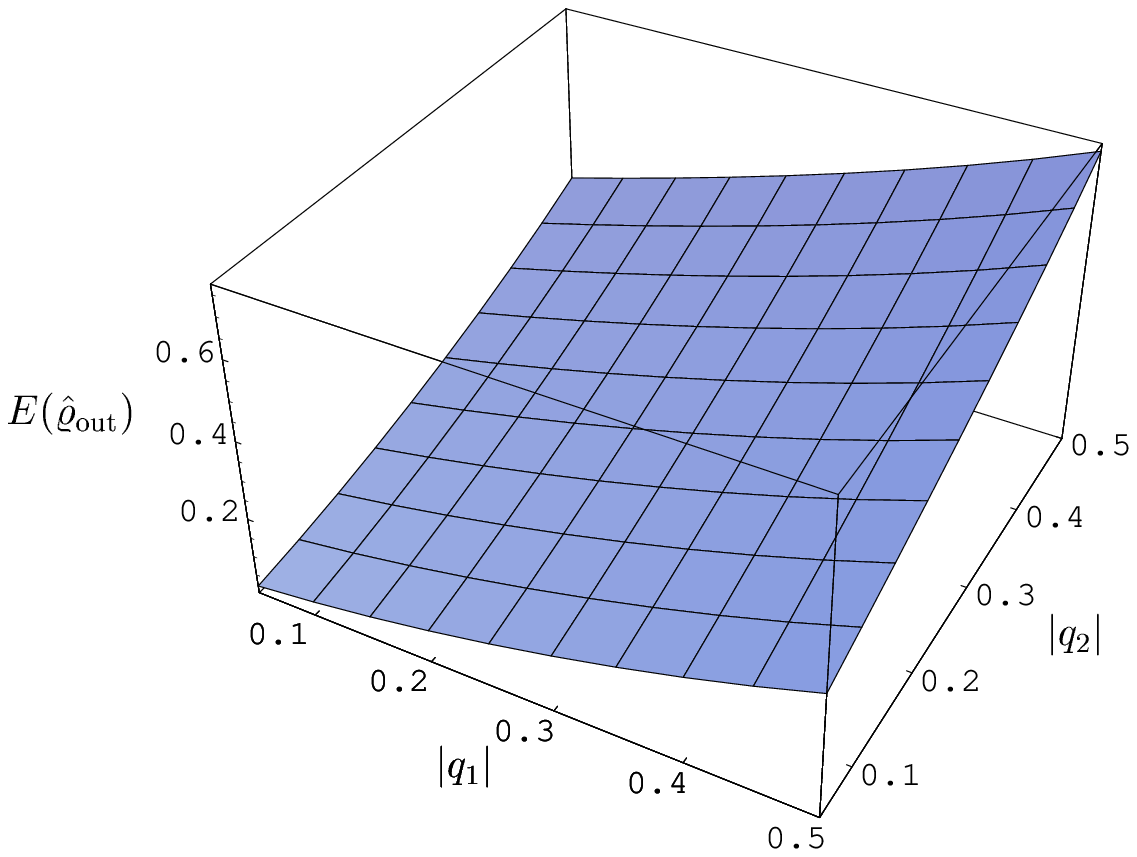,width=7cm}
\caption{\label{maxphase} Entanglement produced at a lossless
$50\%/50\%$ beam splitter by mixing two modes prepared
in squeezed vacuum states
as a function of $|q_1|$ and $|q_2|$ for the phase condition 
$2(\phi_R$ $\!-$ $\!\phi_T)$ $\!+$ $\!\phi_2$ $\!-\phi_1$
$\!=$ $\!\pi$.
}
\end{figure}

\begin{figure}[h]
\psfig{file=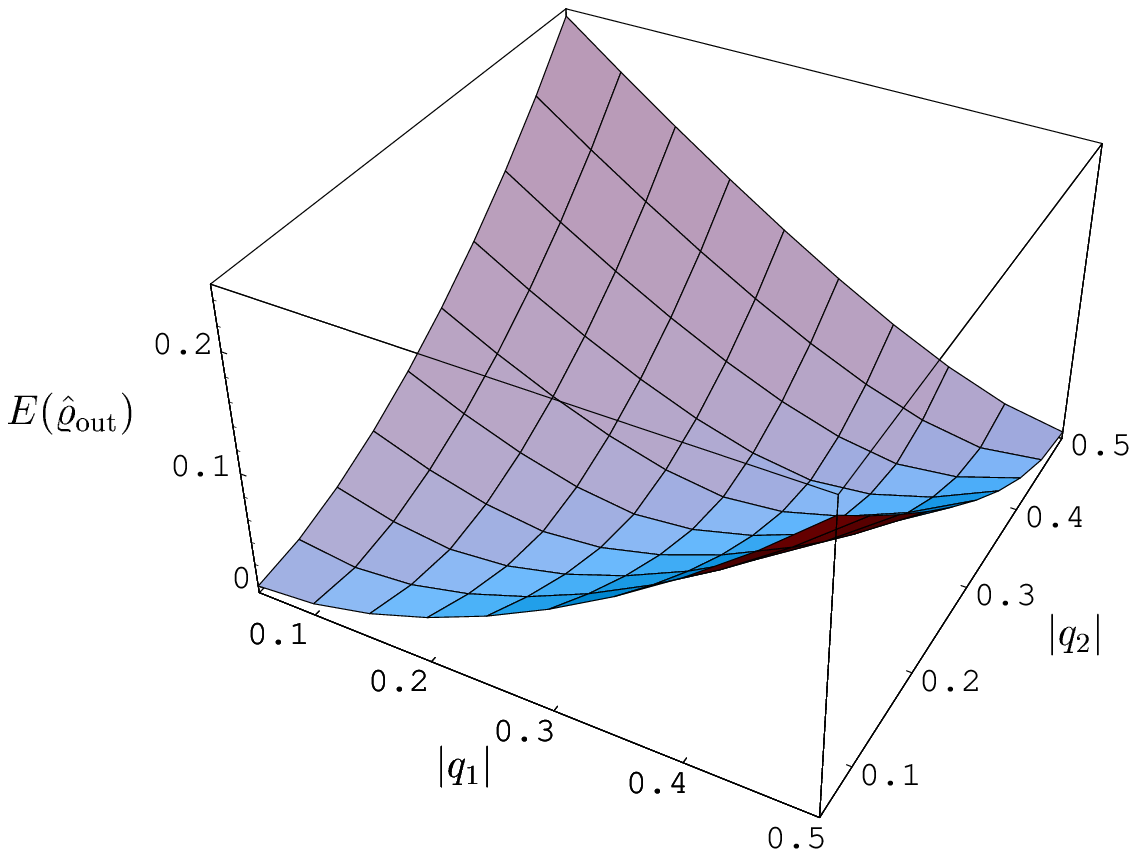,width=7cm}
\caption{\label{minphase} Entanglement produced at a lossless
$50\%/50\%$ beam splitter by mixing two modes prepared
in squeezed vacuum states
as a function of $|q_1|$ and $|q_2|$ for the phase condition
$2(\phi_R$ $\!-$ $\!\phi_T)$ $\!+$ $\!\phi_2$ $\!-\phi_1$
$\!=$ $\!0$.
}
\end{figure}

Since, according to Eq.~(\ref{2.6}), the output quantum state is a
pure state, entanglement is uniquely measured by the von
Neumann entropy of the (reduced) quantum state of either of
the output modes, 
\begin{equation}
\label{2.10}
E(\hat{\varrho}_{\rm out})
   = S_{1(2)} = -{\rm Tr}\bigl[ \hat{\varrho}_{1(2)} \ln
\hat{\varrho}_{1(2)} \bigr],
\end{equation}
where $\hat{\varrho}_{1(2)}$ denotes the (reduced) output
density operator of mode $1(2)$, which is obtained by
tracing $\hat{\varrho}_{\rm out}$
with respect to mode $2(1)$. The result of the
numerical calculation is illustrated in Figs.~\ref{maxphase} and
\ref{minphase} for a $50\%/50\%$ beam splitter. 
In Fig.~\ref{maxphase} the phases are chosen such that the relation
\mbox{$\phi_1$ $\!-$ $\!\phi_2$ $\!+2(\phi_T$ $\!-\phi_R)$ $\!=$
$\!\pm\pi$} is valid, thus leading to a TMSV state for
$|q_1|$ $\!=$ $\!|q_2|$.
Figure \ref{minphase} shows the case where
\mbox{$\phi_1$ $\!-$ $\!\phi_2$
$\!+2(\phi_T$ $\!-\phi_R)$ $\!=$ $0$}, so that for
$|q_1|$ $\!=$ $\!|q_2|$
no entanglement is observed.
Note that using squeezed coherent states instead of squeezed
vacuum states does not change the entanglement.
This is due to the fact that coherent shifts are unitary
operations on subsystems which leave any entanglement measure invariant.

%%%%%%%%%%%%%%%%%%%%%%%%%%%%%%%%%%%%%%%%%%%%%%%%%%%%%%%%%%%%%%%%%%%%%%
\subsection{Lossy beam splitters}
\label{sec:genloss}

In practice there are always some losses and
things get slightly more complicated. The SU(2) group transformation in
Eq.~(\ref{2.1}) has to be replaced by a
SU(4) group transformation, where
the unitary transformation acts in the 
product Hilbert space of the field modes and the
device modes \cite{Knoll99,Scheel00c,Buch}.
As a result, Eqs.~(\ref{2.1}) and (\ref{2.2}), respectively,
have to be replaced by
\begin{equation}
\label{2.20a}
\hat{\mbb{\beta}}(\omega)
   = \mbb{\Lambda}(\omega) \, \hat{\mbb{\alpha}}(\omega)
\end{equation}
and
\begin{equation}
\label{2.20}
\hat{\varrho}_{\rm out}^{({\rm F})} = {\rm Tr}^{({\rm D})}
\hat{\varrho}_{\rm in}\!\left[ \mbb{\Lambda}^+(\omega)
\hat{\mbb{\alpha}}(\omega) , \mbb{\Lambda}^{\rm T}(\omega)
\hat{\mbb{\alpha}}^\dagger(\omega) \right], 
\end{equation}
where the ``four-vector'' notation
$\hat{\mbb{\alpha}}(\omega)$ for abbreviating the list of operators
$\hat{a}_1(\omega)$, $\hat{a}_2(\omega)$, $\hat{g}_1(\omega)$,
$\hat{g}_2(\omega)$ [and $\hat{\mbb{\beta}}(\omega)$ accordingly]
has been used.
The SU(4) group element $\mbb{\Lambda}(\omega)$ is expressed
in terms of the characteristic transformation and absorption
matrices ${\bf T}(\omega)$ and ${\bf A}(\omega)$ of the beam
splitter as
\begin{equation}
\label{2.21}
\mbb{\Lambda}(\omega) = \left(\! \begin{array}{cc}
{\bf T}(\omega) & {\bf A}(\omega) \\[.5ex]
-{\bf S}(\omega) {\bf C}^{-1}(\omega) {\bf T}(\omega) &
{\bf C}(\omega) {\bf S}^{-1}(\omega) {\bf A}(\omega)
\end{array} \!\right)
\end{equation}
with the commuting positive Hermitian matrices
\begin{equation}
\label{2.22}
{\bf C}(\omega)= \sqrt{{\bf T}(\omega){\bf T}^+(\omega)}
\,, \quad
{\bf S}(\omega)= \sqrt{{\bf A}(\omega){\bf A}^+(\omega)} \,.
\end{equation}
{F}rom the above, the output density matrix in the Fock basis
can be given in the form  of (Appendix \ref{app:a}):
\begin{eqnarray}
\label{2.23}
\lefteqn{
\langle m_1,m_2 | \hat{\varrho}_{\rm out}^{({\rm F})} | n_1,n_2
\rangle
}
\nonumber \\[.5ex] &&\hspace{2ex}
= \sqrt{\frac{(1\!-\!|q_1|^2)(1\!-\!|q_2|^2)}{m_1!m_2!n_1!n_2!}}
(-1)^{m_1+m_2+n_1+n_2} 
\nonumber \\[.5ex] &&\hspace{2ex} \times
\sum\limits_{g_1,g_2=0}^\infty \frac{1}{g_1!g_2!}
{\rm H}^{\bf M}_{m_1,m_2,g_1,g_2}({\bf 0})
{\rm H}^{\ast{\bf M}}_{n_1,n_2,g_1,g_2}({\bf 0}),
\end{eqnarray}
where ${\rm H}^{\bf M}_n({\bf 0})$ denotes the Hermite polynomial of four
variables with zero argument, generated by the symmetric matrix
${\bf M}$ with elements
\begin{equation}
\label{2.24}
M_{ij} = q_1 \Lambda_{i1} \Lambda_{j1} +  q_2 \Lambda_{i2}
\Lambda_{j2} \,.
\end{equation}
Note that in Eq.~(\ref{2.23}) it is assumed that the device
is prepared in the ground state.

In order to quantify the entanglement content
of a mixed state $\hat{\varrho}$, such as
$\hat{\varrho}_{\rm out}^{({\rm F})}$ in Eq.~(\ref{2.23}),
we make use of the relative entropy measuring the distance of
the state to the set ${\cal S}$ of all separable states $\hat{\sigma}$
\cite{Vedral98},
\begin{equation}
\label{2.25}
E(\hat{\varrho})= \min_{\hat{\sigma}\in {\cal S}}
{\rm Tr}\big[ \hat{\varrho}
\big( \ln \hat{\varrho} - \ln \hat{\sigma} \big) \big] \,.
\end{equation}
For pure states this measure reduces to the von
Neumann entropy (\ref{2.10}) of either of the subsystems
which can be computed by means of Schmidt decomposition of
the continuous-variable state
\cite{Parker00}.
It is also known that when the quantum state has the Schmidt form, 
\begin{equation}
\label{2.27}
\hat{\varrho} = \sum\limits_{n,m} C_{n,m}
|\phi_n,\psi_n\rangle\langle \phi_m,\psi_m|,
\end{equation}
then the amount of entanglement measured by the
relative entropy is given by \cite{Rains99,Wu00}
\begin{equation}
\label{2.28}
E(\hat{\varrho})
= -\sum_n C_{n,n} \ln C_{n,n} - S(\hat{\varrho}).
\end{equation}
Unfortunately, there is no closed solution of
Eq.~(\ref{2.25}) for arbitrary mixed states. Nevertheless,
upper bounds on the entanglement can be calculated
\cite{Hiroshima00}, representing the quantum state under study
in terms of states in Schmidt decomposition
and using the convexity of the relative entropy,
\begin{equation}
\label{2.26}
E\Bigl(\sum_n p_n\hat{\varrho}_n\Bigr)
\le \sum_n p_n
E(\hat{\varrho}_n),
\quad \sum_n p_n =1.
\end{equation} 

\begin{figure}[h]
\psfig{file=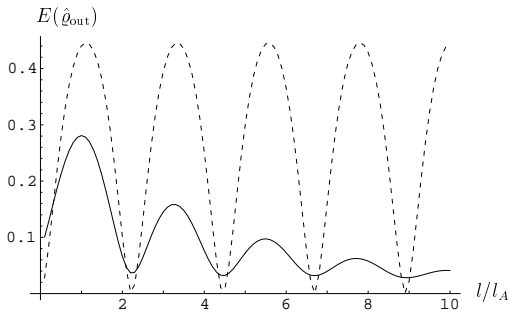,width=7cm}
\caption{\label{lossy}
Entanglement produced at a lossless beam splitter
with refractive index $n$ $\!=$ $\!1.41$ (dashed curve) as a function
of the beam splitter thickness $l$. The full curve shows the upper
bound of the produced entanglement at a lossy beam splitter with
\mbox{$n$ $\!=$ $\!1.41+0.1i$}. The squeezing parameters chosen
are \mbox{$q_1$ $\!=$ $\!q_2$ $\!=$ $\!0.5$}.}
\end{figure}

Applying the method to the output quantum state in
Eq.~(\ref{2.23}), i.e., rewriting it in the form of
\begin{eqnarray}
\label{2.29}
\hat{\varrho}_{\rm out}^{({\rm F})}
&=& \sum\limits_{k,l=0}^\infty
   C_{k,l,0} |k,k\rangle\langle l,l|
\nonumber \\ &&
   +\sum\limits_{m=1}^\infty \sum\limits_{k,l=0}^\infty C_{k,l,m}
   |k+m,k\rangle\langle l+m,l|
\nonumber \\ &&
   +\sum\limits_{m=1}^\infty \sum\limits_{k,l=0}^\infty C_{k,l,m}
   |k,k+m\rangle\langle l,l+m|
\nonumber \\ 
&=& p_0 \hat{\varrho}_0
+ \sum\limits_{m=1}^\infty p_m \hat{\varrho}_{m,1}
+ \sum\limits_{m=1}^\infty p_m \hat{\varrho}_{m,2}\,,
\end{eqnarray}
the inequality (\ref{2.26}) leads to
\begin{equation}
\label{2.30}
E\big(\hat{\varrho}_{\rm out}^{({\rm F})}\big)
\le p_0
E(\hat{\varrho}_0)
+\sum\limits_{m=1}^\infty p_m \left[
E(\hat{\varrho}_{m,1})
+E(\hat{\varrho}_{m,2})
\right],
\end{equation}
where $E(\hat{\varrho}_0)$, $E(\hat{\varrho}_{m,1})$, and
$E(\hat{\varrho}_{m,2})$ can be determined according to
Eq.~(\ref{2.28}). In the numerical calculation we have
used the dielectric-plate model of a beam splitter,
taking the ${\bf T}$ and ${\bf A}$ matrices from 
\cite{Buch,Gruner96}. The result is illustrated in
Fig.~\ref{lossy}, which shows the dependence on the
plate thickness of the upper bound of the attainable
entanglement. The oscillations are due to phase matching
and phase mismatch at certain beam splitter thicknesses
[cf. Eq.~(\ref{2.7})]. Note that the local minima of the curve
for the lossy beam splitter never go down to zero as do the
corresponding minima of the curve for the lossless beam splitter.
This obviously reflects the fact that the result for the
lossless beam splitter is exact, whereas that
for the lossy beam splitter is only an upper bound.

%%%%%%%%%%%%%%%%%%%%%%%%%%%%%%%%%%%%%%%%%%%%%%%%%%%%%%%%%%%%%%%%%%%%%%
\section{Entanglement degradation in TMSV transmission through lossy
optical fibers}
\label{sec:deg}

\begin{figure}[h]
\psfig{file=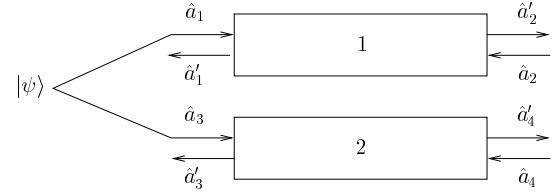,width=7cm}
\caption{\label{eightport} A two-mode input field prepared in the
quantum state \protect$|\psi\rangle$ is transmitted through two
absorbing dielectric four-port devices, \protect$\hat{a}_1$,
$\hat{a}_3$ ($\hat{a}_2'$, $\hat{a}_4'$) being the photonic operators
of the relevant input (output) modes.}
\end{figure}

Let us now turn to the problem of entanglement degradation
in transmission of light prepared in a TMSV state through absorbing
fibers. The situation is somewhat different from that in 
the previous section, since we are effectively dealing with an
eight-port device as depicted in Fig.~\ref{eightport}, where
the two channels are characterized by the transmission
($T_i$) and reflection ($R_i$) coefficients (\mbox{$i$ $\!=$ $\!1,2$}).
In particular for perfect input coupling (\mbox{$R_i$ $\!=$ $\!0$}),
the system is essentially characterized by the transmission
coefficients $T_i$.

{F}rom Eq.~(\ref{2.8a}) it is easily seen that in
the Fock basis a TMSV state reads
\begin{equation}
\label{3.1}
|\mbox{TMSV}\rangle
= \sqrt{1-|q|^2} \,\sum\limits_{n=0}^\infty (-q)^n |n,n\rangle,
\end{equation}
whose entanglement content is
\begin{equation}
\label{3.12}
E\big(|\mbox{TMSV}\rangle\big) =
-\ln \big( 1\!-\!|q|^2 \big) -\frac{|q|^2}{1\!-\!|q|^2} \ln |q|^2 .
\end{equation}
Application of the quantum-state transformation (\ref{2.20})
yields (\mbox{$R_i$ $\!=$ $\!0$}) \cite{Scheel00c}
\begin{eqnarray}
\label{3.2}
\lefteqn{
\hat{\varrho}_{\rm out}^{({\rm F})} = ( 1-|q|^2)
\sum\limits_{m=0}^\infty \sum\limits_{k,l=0}^\infty
\Big[ K_{k,l,m}
}
\nonumber \\ &&\hspace{2ex} \times\,
( c_m |m\!+\!k\rangle\langle k| +\mbox{H.c.})
\otimes
( d_m |m\!+\!l\rangle\langle l| +\mbox{H.c.})\Big],
\end{eqnarray}
where
\begin{eqnarray}
\label{3.3}
c_m &=& (-q)^{m/2} T_1^m
\left( 1-{\textstyle\frac{1}{2}}\delta_{m0} \right), \\
d_m &=& (-q)^{m/2} T_2^m
\left( 1-{\textstyle\frac{1}{2}}\delta_{m0} \right),
\end{eqnarray}
and
\begin{eqnarray}
\label{3.4}
K_{k,l,m} &=&
\frac{\big[|q|^2 (1\!-\!|T_1|^2)
(1\!-\!|T_2|^2) \big]^a a! (a\!+\!m)!}
{\sqrt{k!l!(k\!+\!m)!(l\!+\!m)!}(a\!-\!k)!(a\!-\!l)!}
\nonumber \\ && \hspace{-7ex}
\times
\left( \frac{|T_1|^2}{1-|T_1|^2} \right)^k
\left( \frac{|T_2|^2}{1-|T_2|^2} \right)^l
\nonumber \\ && \hspace{-7ex}
\times\,
{}_2F_1\!\left[ {a\!+\!1,\,a\!+\!m\!+\!1
\atop |k\!-\!l|\!+\!1} \,;\,
|q|^2(1\!-\!|T_1|^2)(1\!-\!|T_2|^2) \right]
\nonumber \\
\end{eqnarray}
[$a$ $\!=$ $\!\max(k,l)$]. Note that in Eq.~(\ref{3.2}) the fibers
are assumed to be in the ground state.
 
%%%%%%%%%%%%%%%%%%%%%%%%%%%%%%%%%%%%%%%%%%%%%%%%%%%%%%%%%%%%%%%%%%%%%%
\subsection{Entanglement estimate by pure state extraction}

The amount of entanglement contained in the (mixed)
output state (\ref{3.2}) can also be estimated, 
following the line sketched in Section \ref{sec:genloss}.
In particular, the convexity of the relative entropy
can be combined with Schmidt decompositions of the
output state in order to calculate, on using the theorem
(\ref{2.28}), appropriate bounds on entanglement.
Before doing so, let us first consider the
case of low initial squeezing, for which the 
entanglement can be estimated rather simply.

%%%%%%%%%%%%%%%%%%%%%%%%%%%%%%%%%%%%%%%%%%%%%%%%%%%%%%%%%%%%%%%%%%%%%%
\subsubsection{Extraction of a single pure state}
\label{sec:single}

\begin{figure}[h]
\psfig{file=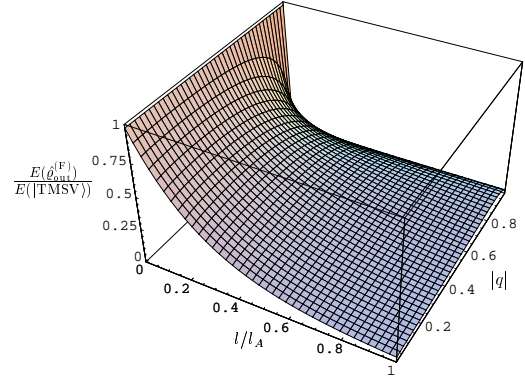,width=7cm}
\caption{\label{estimate}
Estimate of the entanglement, Eq.~(\protect\ref{3.15}),
observed after transmission of a TMSV through
absorbing fibers \mbox{($T_1$ $\!=$ $\!T_2$)}    
as a function of the squeezing parameter $|q|^2$
and the transmission length $l$.}
\end{figure}

Since, by Eqs.~(\ref{3.2}) --
(\ref{3.4}), for low squeezing only a few matrix elements are excited
which were not contained in the original Fock expansion (\ref{3.1}),
we can forget about the entanglement that could be present in the newly
excited elements and treat them as contributions to the separable
states only. Following \cite{Scheel00c}, the inseparable
state relevant for entanglement can then be estimated to
be the pure state  
\begin{equation}
\label{3.13}
\sqrt{1-\lambda}\, |\Psi\rangle 
= \sqrt{\frac{1-|q|^2}{K_{000}}} \sum\limits_{n=0}^\infty
K_{00n} c_n d_n |n,n\rangle.
\end{equation}
It has the properties that only matrix elements 
of the same type as in the input TMSV state occur and
the coefficients of the matrix elements \mbox{$|0,0\rangle$ 
$\!\leftrightarrow$ $\!|n,n\rangle$} are met exactly, i.e.,
\begin{equation}
\label{3.14}
(1-\lambda)\langle 0,0|\Psi\rangle\langle\Psi|n,n\rangle
=\langle 0,0|\hat{\varrho}_{\rm out}^{\rm (F)}|n,n\rangle. 
\end{equation} 
In this approximation, the calculation of the entanglement
of the mixed output quantum state 
reduces to the determination of the entanglement of a
pure state \cite{Scheel00c}:
\begin{eqnarray}
\label{3.15}
\lefteqn{
E(\hat{\varrho}_{\rm out}^{({\rm F})}) 
\approx (1-\lambda)\,E(|\Psi\rangle) 
}
\nonumber\\[.5ex]&&\hspace{2ex}
   =\frac{1\!-\!x}{(1\!-\!x)^2\!-\!y}\,
   \ln\!\left[\frac{1\!-\!x}{(1\!-\!x)^2\!-\!y}\right] 
\nonumber\\[.5ex]&&\hspace{2ex}
   +\, \frac{(1\!-\!x)\{[y\!+\!(1\!-\!x)^2] 
   \ln (1\!-\!x)\!-\!y \ln y\}}{[y\!-\!(1\!\!-x)^2]^2} \,,
\end{eqnarray}
where
\begin{equation}
\label{3.16}
x = |q|^2 (1-|T_1|^2) (1-|T_2|^2) ,
\end{equation}
\begin{equation}
\label{3.17}
y = |qT_1T_2|^2 .
\end{equation}
Note that for $T_1$ $\!=$ $\!T_2$ $\!=$ $\!1$ the 
entanglement of the TMSV state is preserved, i.e.,
Eq.~(\ref{3.15}) reduces to Eq.~(\ref{3.12}).
In Fig.~\ref{estimate}, the estimate of entanglement as given by
Eq.~(\ref{3.15}) is plotted as a function of the
transmission length and the strength of initial
squeezing for $T_1$ $\!=$ $\!T_2$ $\!=$ $\!T$, where $T$ is given by
the Lambert--Beer law of extinction,
\begin{equation}
\label{3.11}
T = {\rm e}^{in_{\rm R}(\omega)\omega l/c} {\rm e}^{-l/l_{\rm A}}.
\end{equation}
Here, $n_{\rm R}$ is the real part of the complex refractive
index, $l_{\rm A}$ $\!=$ $\!c/(n_{\rm I}\omega)$ is the
absorption length, and $l$ is the transmission length.

It is worth repeating that the estimate given by
Eq.~(\ref{3.15}) is valid for low squeezing
only. Higher squeezing amounts to more excited density matrix elements 
and Eq.~(\ref{3.15}) might become wrong.
Moreover, we cannot even infer it to be a {\it bound} in any sense
since no inequality has been involved. A possible way out would be to
extract successively more and more pure states from
Eq.~(\ref{3.2}). But instead, let us turn to the Schmidt
decomposition. 

%%%%%%%%%%%%%%%%%%%%%%%%%%%%%%%%%%%%%%%%%%%%%%%%%%%%%%%%%%%%%%%%%%%%%%
\subsubsection{Upper bound of entanglement}
\label{sec:rains}

\begin{figure}[h]
\psfig{file=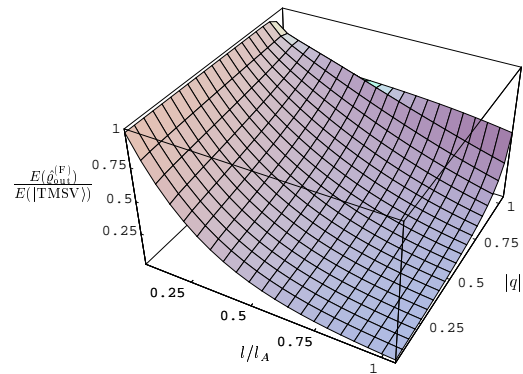,width=7cm}
\caption{\label{tmsv_est}
Upper bound on the entanglement degradation of a TMSV
transmitted through absorbing
fibers \mbox{($T_1$ $\!=$ $\!T_2$)} as a function of the
squeezing parameter $|q|$
and the transmission length $l$.
In the numerical calculation, at most 30 photons per mode
have been taken into account which is obviously not sufficient
for higher squeezing when higher photon-number states are
excited.
}
\end{figure}

In a similar way as in Section \ref{sec:genloss},
an upper bound on the entanglement can be obtained \cite{Hiroshima00},
if the density operator (\ref{3.2})
is rewritten as the convex sum of density operators in Schmidt
decomposition,
\begin{eqnarray}
\label{3.20}
\hat{\varrho}_{\rm out}^{({\rm F})} &=& \sum\limits_{k,l=0}^\infty \bigg\{
A_{k,l} |k,k\rangle\langle l,l| \nonumber \\ &&
+\sum\limits_{m=1}^\infty B_{k,l,m} |k+m,k\rangle\langle l+m,l|
\nonumber \\ &&
+\sum\limits_{m=1}^\infty C_{k,l,m} |k,k+m\rangle\langle l,l+m|
\bigg\},
\end{eqnarray}
and the inequality (\ref{2.26}) together with Eq.~(\ref{2.28})
is applied.
The result is illustrated in Fig.~\ref{tmsv_est}.

{F}rom general arguments one would expect the entanglement to decrease
faster the more squeezing one puts into the TMSV, because stronger
squeezing is equivalent to saying the state is more macroscopically
non-classical and quantum correlations should be destroyed faster. As
an example, one would have to look at the entanglement degradation of
an $n$-photon Bell-type state $|\Psi^\pm_n\rangle$, 
$E(|\Psi^\pm_n\rangle)$ $\!\le$ $\!|T|^{2n} \ln 2$ \cite{Scheel00a}.
Since the transmission coefficient $T$ decreases exponentially
with the transmission length, entanglement decreases even faster.
Note that similar arguments also hold for the destruction of
the interference pattern of a cat-like state
$\sim|\alpha\rangle+|-\alpha\rangle$ when it is transmitted,
e.g., through a beam splitter. It is well known 
that the two peaks (in the $j$th output channel) decay as $|T_{j1}|^2$,
whereas the quantum interference decays as
$|T_{j1}|^2 \exp[-2|\alpha|^2(1$ $\!-$ $\!|T_{j1}|^2)]$. 

\begin{figure}[h]
\psfig{file=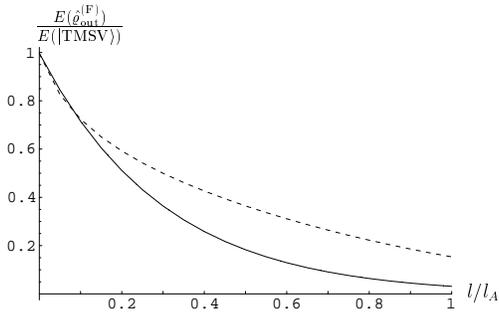,width=7cm}
\caption{\label{vergleich} Upper bound of the entanglement
degradation of a TMSV transmitted through absorbing
fibers \mbox{($T_1$ $\!=$ $\!T_2$)} as a function of the
transmission length $l$ for the squeezing parameters
\mbox{$q$ $\!=$ $\!0.1$} (solid line)
and \mbox{$q$ $\!=$ $\!0.9$} (dashed line).
In the numerical calculation, Fock states $|n\rangle$ up to
\mbox{$q^n$ $\!\lesssim$ $\!0.02$} have been taken into account.
}
\end{figure}

The upper bound on the entanglement as calculated above
seems to suggest that the entanglement degradation
is simply exponential with the transmission length for
essentially all (initial) squeezing parameters, which would
make the TMSV a good candidate for a robust entangled quantum
state. But this is a fallacy. The higher the
squeezing, the more density matrix elements are excited, and the more
terms appear, according to Eq.~(\ref{3.20}), in the
convex sum (\ref{2.26}). Equivalently, more and
more separable states are mixed into the full quantum state.
By that, the inequality gets more inadequate. In order to see this
better, we have shown in Fig.~\ref{vergleich} the upper bound on the
entanglement for just two different (initial) squeezing
parameters $|q|$ $\!=$ $\!0.71$ (equivalent to the mean photon number
of $\bar{n}$ $\!=$ $\!1$, solid line) and $|q|$ $\!=$ $\!0.9535$
(\mbox{$\bar{n}$ $\!=$ $\!10$}, dashed line). 
For small transmission lengths, hence very few separable states are
mixed in, the curves show the expected behavior in the sense that the 
state with higher initial squeezing decoheres fastest. The behavior
changes for larger transmission lengths. We would thus conclude that
the upper bound proposed in \cite{Hiroshima00} is insufficient.

%%%%%%%%%%%%%%%%%%%%%%%%%%%%%%%%%%%%%%%%%%%%%%%%%%%%%%%%%%%%%%%%%%%%%%
\subsection{Distance to separable Gaussian states}
\label{sec:distance}

The methods of computing entanglement estimates and bounds as
considered in the preceding sections are  based on Fock-state
expansions. In practice they are typically restricted to situations
where only a few quanta of the overall system (consisting of
the field and the device) are excited, 
otherwise the calculation even of the matrix elements
becomes arduous. Here we will focus on another way
of computing the relative entropy, which will also enable us to
give an essentially better bound on the entanglement
(for other quantities that characterize, in a sense,
entanglement, see \cite{Chizhov01}).  

Since it is close to impossible to compute the distance of a Gaussian
state to the set of {\em all} separable states we restrict ourselves
to separable {\em Gaussian} states. A quantum state is commonly called
Gaussian if its quantum characteristic function is Gaussian. By the
general relation for a $N$-mode quantum state
\begin{equation}
\label{4.1}
\hat{\sigma} = \frac{1}{\pi^N} \int {\rm d}^{2N}\!\mbb{\alpha}\,
\chi(-\mbb{\alpha}) \hat{D}(\mbb{\alpha})
\end{equation}
it is obvious that the density operator of a Gaussian state
can be written in exponential form of
\begin{equation}
\label{4.2}
\hat{\sigma} = {\cal N}
\exp\!\left[ -\left(\hat{a}^\dagger \, \hat{a}\right) {\bf M}_\sigma
{\hat{a} \choose \hat{a}^\dagger } \right],
\end{equation}
where ${\bf M}_\sigma$ is a Hermitian matrix that can be assumed
to give a symmetrically ordered density operator, and ${\cal N}$
is a suitable normalization factor. Here and in the following
we restrict ourselves to Gaussian states with zero mean.
Since coherent displacements, being local unitary transformations,
do not influence entanglement, they can be disregarded. 

The relative entropy (\ref{2.25}) can now be written as
\begin{eqnarray}
\label{4.3}
\lefteqn{
E_R\big(\hat{\varrho}\big)
   = \min_{\hat{\sigma}\in{\cal S}} {\rm Tr}
   \left\{
   \hat{\varrho}
   \left[
   \ln \hat{\varrho} - \ln {\cal N}
   +\left(\hat{a}^\dagger \, \hat{a}\right) {\bf M}_\sigma
   {\hat{a} \choose \hat{a}^\dagger }
   \right]
   \right\}
}
\nonumber \\&&\hspace{2ex}
   = {\rm Tr}\left(\hat{\varrho} \ln \hat{\varrho}\right)
   + \min_{\hat{\sigma}\in{\cal S}}
   \left\langle
   \left(\hat{a}^\dagger \, \hat{a}\right) {\bf M}_\sigma
   {\hat{a} \choose \hat{a}^\dagger } -\ln {\cal N}
   \right\rangle_{\!\hat{\varrho}}.
\end{eqnarray}
Since we have chosen the density operator $\hat{\sigma}$ to be
symmetrically ordered, the last term in Eq.~(\ref{4.3}) is nothing but 
a sum of (weighted) symmetrically ordered expectation values
$\langle \hat{a}^{\dagger m} \hat{a}^n \rangle_{s=0}$
($m$ $\!+$ $\!n$ $\!=$ $\!2$). For a Gaussian quantum state
$\hat{\varrho}$ it can be shown (Appendix~\ref{app:b})
that Eq.~(\ref{4.3}) can equivalently be written in terms of
the matrix ${\bf D}_\varrho$ in the exponential of the
characteristic function of $\hat{\varrho}$ as
\begin{equation}
\label{4.4}
E_R\big(\hat{\varrho}\big) = {\rm Tr}\left( \hat{\varrho} \ln
   \hat{\varrho} \right)
   + \min_{\hat{\sigma}\in{\cal S}}
   \!\left[ \textstyle\frac{1}{2}
   {\rm Tr}\left({\bf M}_\sigma {\bf D}_\varrho\right)
   -\ln {\cal N} \right] . 
\end{equation}

{F}rom the above it is clear that we only need the matrix
${\bf D}_\varrho$ (which is unitarily equivalent to the variance
matrix). For a Gaussian distribution with zero mean the 
elements of the variance matrix ${\bf V}$ are defined by
\mbox{$V_{ij}$ $\!=$
$\!\langle {\hat{\zeta}_i,\hat{\zeta}_j } \rangle_{s=0}$} as the
(symmetrically ordered) expectation values of the quadrature
components \mbox{$\hat{\mbb{\zeta}}$ $\!=$
$\!(\hat{x}_1,\hat{p}_1,\hat{x}_2,\hat{p}_2)$}.

The variance matrix of the TMSV state (\ref{3.1}) reads
(\mbox{$q$ $\!=$ $\!\tanh|\xi|e^{i\phi}$},
\mbox{$\xi$ $\!=$ $\!|\xi|e^{i\phi}$})
\begin{equation}
\label{4.6}
{\bf V}_{\varrho} = \left(
\begin{array}{cc} {\bf X} & {\bf Z} \\ {\bf Z}^{\rm T} & {\bf Y} \end{array}
\right) = \left(
\begin{array}{cccc}
c/2&0&-s_1/2&-s_2/2\\0&c/2&-s_2/2&s_1/2\\
-s_1/2&-s_2/2&c/2&0\\-s_2/2&s_1/2&0&c/2
\end{array} \right)\!,
\end{equation}
with the notation \mbox{$c$ $\!=$ $\!\cosh 2|\xi|$},
\mbox{$s_1$ $\!=$ $\!\sinh 2|\xi|\cos\phi$}, and
\mbox{$s_2$ $\!=$ $\!\sinh 2|\xi|\sin\phi$}.
In case \mbox{$\phi$ $\!=$ $\!0$} the variance matrix (\ref{4.6})
reduces to the generic form
\begin{equation}
\label{4.6a}
{\bf V}_0 = \left(
\begin{array}{cccc}
x&0&z_1&0\\0&x&0&z_2\\z_1&0&y&0\\0&z_2&0&y
\end{array} \right)
\end{equation}
specified by four real parameters. Note that the variance matrix of
any Gaussian state can be brought to the form (\ref{4.6a}) by local
Sp(2,$\mathbb{R}$)$\otimes$Sp(2,$\mathbb{R}$) transformations
\cite{Simon00}, so that we can restrict further discussions to that case.

Application of the input-output relations (\ref{2.20a})
gives for the elements of the variance matrix of the output state, 
on assuming that the two modes are transmitted through
two four-port devices prepared in thermal states of mean photon
numbers $n_{{\rm th}i}$,
\cite{Scheel00c}
\begin{eqnarray}
\label{4.7}
\lefteqn{
X_{11} = X_{22} =
}
\nonumber \\ &&\hspace{2ex}
\textstyle\frac{1}{2} c|T_1|^2
+\textstyle\frac{1}{2} |R_1|^2 +\left( n_{{\rm th}1}
\!+\!\textstyle\frac{1}{2} \right) \left( 1\!-\!|T_1|^2\!-\!|R_1|^2 \right),
\\
\lefteqn{
Y_{11} = Y_{22} =
} \nonumber \\ &&\hspace{2ex}
\textstyle\frac{1}{2} c|T_2|^2
+\textstyle\frac{1}{2} |R_2|^2 +\left( n_{{\rm th}2}
\!+\!\textstyle\frac{1}{2} \right) \left( 1\!-\!|T_2|^2-|R_2|^2 \right),
\end{eqnarray}
\begin{eqnarray}
\label{4.8}
Z_{11} = -Z_{22} &=& -\textstyle\frac{1}{2} s\,
   {\rm Re}\left(T_1T_2\right),
\\
Z_{12} = Z_{21} &=& -\textstyle\frac{1}{2} s\,
   {\rm Im}\left(T_1T_2\right)
\end{eqnarray}
($\phi$ $\!=$ $\!0$).
With regard to optical fibers with perfect input coupling
(\mbox{$R_i$ $\!=$ $\!0$}) and equal transmission lengths,
we again may set \mbox{$|T_i|$ $\!=$ $\!e^{-l/l_{\rm A}}$}. Moreover,
we may assume real $T_i$ and thus set $Z_{12}$ $\!=$ $\!Z_{21}$
$\!=$ $\!0$.

First, one can check for separability according to the criterion
\cite{Duan00a,Simon00}
\begin{eqnarray}
\label{4.10}
&& \det{\bf X} \det{\bf Y}+
\left(\textstyle\frac{1}{4}-|\det{\bf Z}|\right)^2
-{\rm Tr}\left({\bf XJZJYJZ}^{\rm T}{\bf J}\right)
\nonumber \\ &&\hspace{2ex}
\ge \textstyle\frac{1}{4}
\left( \det{\bf X} +\det{\bf Y} \right) \,,
\end{eqnarray}
which reduces to 
\begin{equation}
\label{4.11}
4(xy-z_1^2)(xy-z_2^2) \ge (x^2+y^2) +2|z_1z_2|
   -\textstyle\frac{1}{4} \,.
\end{equation}
Combining Eqs.~(\ref{4.7}) -- (\ref{4.11}), it is not
difficult to prove that the
boundary between separability and inseparability is
reached for \cite{Duan00a,Scheel00c,Lee00}
\begin{equation}
\label{4.9}
l = l_{\rm S}
   \equiv \frac{l_{\rm A}}{2} \ln\!\left[ 1+ \displaystyle\frac{1}{n_{\rm th}}
\left( 1-e^{-2|\xi|} \right) \right] .
\end{equation}

It is worth noting that this is exactly the same condition as for the
transmitted state still being a squeezed state or not.
To show this, we calculate the normally-ordered variance
$\langle:\!(\Delta\hat{F})^2)\!: \rangle$ of a phase-sensitive
 quantity such as \mbox{$\hat{F}$ $\!=$ $\!|F_1|e^{i\varphi_{1}}
 \hat{a}_1$ $\!+$
$\!|F_2|e^{i\varphi_{2}}\hat{a}_2$ $\!+$ $\!\mbox{H.c.}$}.
Using the input-output relations (\ref{2.20a}), the normally
ordered variance of the output field is derived to be
\begin{eqnarray}
\label{4.12}
\lefteqn{
\langle:\!(\Delta\hat{F})^2)\!: \rangle_{\rm out} =
2|F_1|^2 \left[ |T_1|^2 \sinh^2|\xi| \!+\! n_{{\rm th}1}
\left( 1\!-\!|T_1|^2 \right) \right] }
\nonumber \\ &&\hspace{2ex}
+\,2|F_2|^2 \left[ |T_2|^2 \sinh^2|\xi| +n_{{\rm th}2}
\left( 1-|T_2|^2 \right) \right]
\nonumber \\ &&\hspace{2ex}
-\,2|F_1F_2T_1T_2| \sinh2|\xi| \,
\cos(\varphi_{1}\!+\!\varphi_{2}\!+\!\varphi_T\!+\!\phi)
\end{eqnarray}
[$T_i$ $\!=$ $|T_i|e^{i\varphi_{T_i}}$, $i\!=\!1,2$; $\varphi_T$ $\!=$
$\!\varphi_{T_1}$ $\!+$ $\!\varphi_{T_2}$].
For equal amplitudes \mbox{$|F_1|$ $\!=$ $\!|F_2|$ $\!=$ $\!|F|$} and
equal fibers \mbox{$|T_1|$ $\!=$ $\!|T_2|$ $\!=$ $\!|T|$},
\mbox{$n_{{\rm th}1}$ $\!=$ $\!n_{{\rm th}2}$ $\!=$ $\!n_{\rm th}$} the
(phase-dependent) minimum is obtained to be
\begin{eqnarray}
\label{4.13}
\lefteqn{
\left.\langle:\!(\Delta\hat{F})^2)\!: \rangle_{\rm out}\right|_{\rm min}
}
\nonumber\\&&\hspace{2ex}
= 4|F|^2 \left[ n_{\rm th} \left( 1\!-\!|T|^2 \right)
\!-\!|T|^2 \sinh|\xi| e^{-|\xi|} \right] .
\end{eqnarray}
Equation (\ref{4.13}) exactly leads to the condition (\ref{4.9}), i.e.,
\begin{equation}
\label{4.13a}
\left.\langle :\!(\Delta\hat{F})^2)\!: \rangle_{\rm out}\right|_{\rm min}
\left\{
\begin{array}{l@{\quad{\rm if}\quad}l}
< 0 & l < l_{\rm S}\,, \\[.5ex]
\ge 0 & l \ge l_{\rm S}\,.
\end{array}
\right.
\end{equation}
Therefore, measurement of squeezing corresponds,
in some sense, to an entanglement measurement.

\begin{figure}[h]
\psfig{file=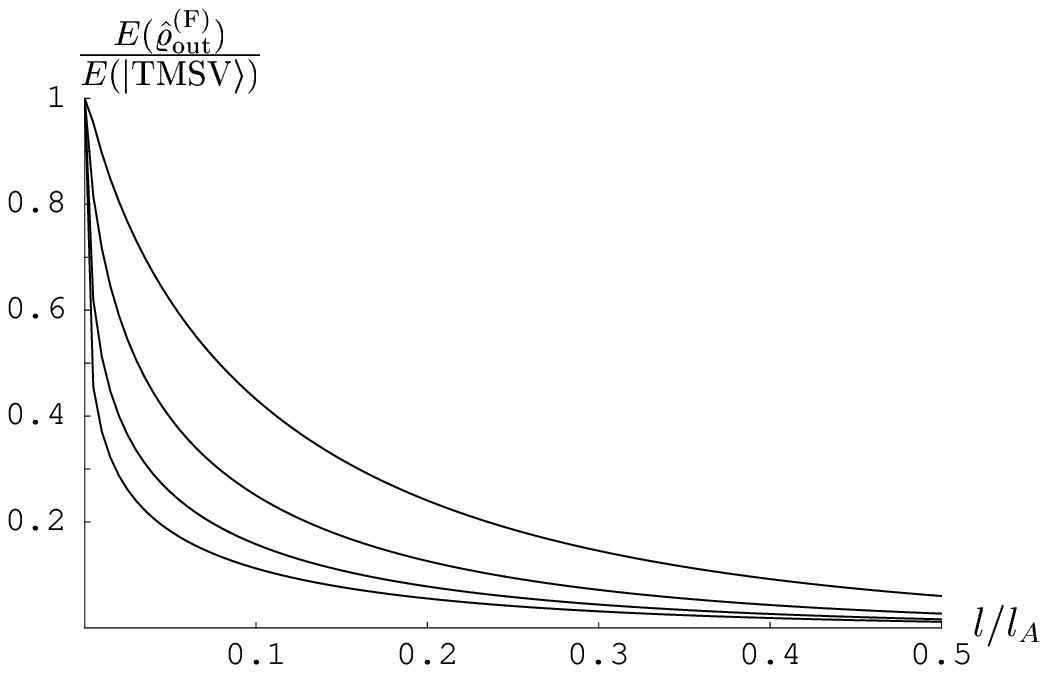,width=8cm}
\caption{\label{distrel}
Entanglement
degradation of a TMSV transmitted through absorbing
fibers \mbox{($T_1$ $\!=$ $\!T_2$)} as a function of the
transmission length $l$ for the (initial) mean photon numbers
\mbox{$\bar{n}$ $\!=$ $1$} (\mbox{$|q|$ $\!\simeq$ $\!0.7071$})
(topmost curve),
\mbox{$\bar{n}$ $\!=$ $10$} (\mbox{$|q|$ $\!\simeq$ $\!0.9535$}),
\mbox{$\bar{n}$ $\!=$ $10^2$} (\mbox{$|q|$ $\!\simeq$ $\!0.9950$}), and
\mbox{$\bar{n}$ $\!=$ $10^3$} (\mbox{$|q|$ $\!\simeq$ $\!0.9995$})
(lowest curve).
}
\end{figure}

In order to obtain (for $l$ $\!<$ $\!l_{\rm S}$) a measure of the
entanglement degradation, we compute the distance of the output
quantum state to the set of all Gaussian states satisfying the
{\em equality} in (\ref{4.11}), since they just represent the
boundary between separability and inseparability. 
These states are completely specified by only three
real parameters [one of the parameters in the equality in (\ref{4.11})
can be computed by the other three]. With regard to Eq.~(\ref{4.4}),
minimization is thus only performed in a three-dimensional
parameter space.
Results of our numerical analysis are shown in
Fig.~\ref{distrel}. It is clearly seen that the entanglement
content (relative to the entanglement in the initial TMSV)
decreases noticeably faster for larger squeezing, or equivalently,
for higher mean photon number [the relation between the
mean photon number $\bar{n}$ and the squeezing parameters
being $\bar{n}$ $\!=$ $\!\sinh^2|\xi|$ $\!=$
$\!|q|^2/(1$ $\!-$ $\!|q|^2)$].

\begin{figure}[h]
\psfig{file=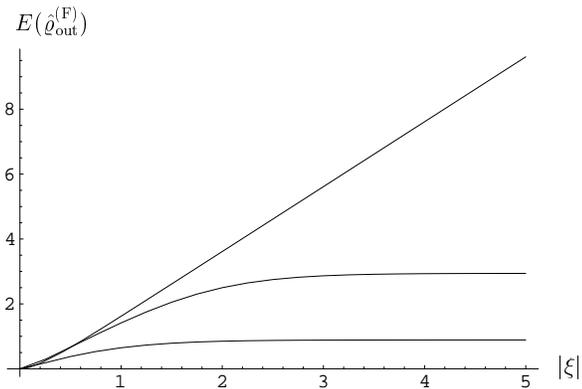,width=8cm}
\caption{\label{laenge}
Available entanglement after transmission of a TMSV
through absorbing fibers \mbox{($T_1$ $\!=$ $\!T_2$)}
as a function of the squeezing parameter $\xi$ for various
transmission lengths $l$ [
\mbox{$l$ $\!=$ $\!0$} (topmost curve),
\mbox{$l$ $\!=$ $\!10^{-2}l_A$} (middle curve),
\mbox{$l$ $\!=$ $\!10^{-1}l_A$} (lowest curve)].
For \mbox{$|\xi|$ $\!\lesssim$ $\!0.5$} and
\mbox{$l/l_A$ $\!\lesssim$ $\!10^{-2}$},
the numerical accuracy of the values of
$E(\hat{\varrho}_{\rm out}^{\rm (F)})$ decreases due to
low accuracy in the eigenvector computation.}
\end{figure}

It is very instructive to know how much entanglement
is available after transmission of the TMSV through the fibers.
Examples of the (maximally) available entanglement 
for different transmission lengths are shown in Fig.~\ref{laenge}.
One observes that a chosen transmission length
allows only for transport of a certain amount of
entanglement. The saturation value, which is quite independent
of the value of the input entanglement,
drastically decrease with increasing transmission length
(compare the upper curve with the two lower curves in the figure).   
This has dramatic consequences for applications in quantum information
processing such as continuous-variable teleportation, where
a highly squeezed TMSV is required in order
to teleport an arbitrary quantum state with sufficiently
high fidelity \cite{Braunstein98}.
Even if the input TMSV would be infinitely squeezed,
the available (low) saturation value of entanglement principally
prevents one from high-fidelity teleportation of {\em arbitrary}
quantum states over finite distances.

%%%%%%%%%%%%%%%%%%%%%%%%%%%%%%%%%%%%%%%%%%%%%%%%%%%%%%%%%%%%%%%%%%%%%%
\subsection{Comparison of the methods}
\label{sec:comparison}

\begin{figure}[h]
\psfig{file=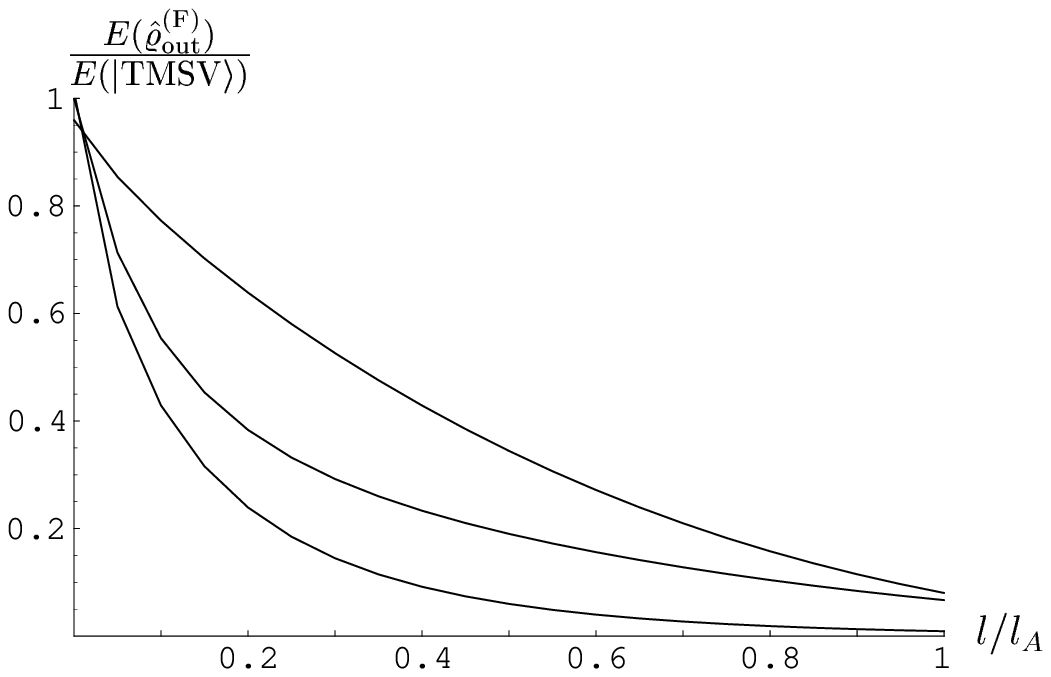,width=8cm}
\caption{\label{abstandsvergleich} Comparison of the upper bound on
entanglement (upper curve) according to Fig.~\protect\ref{tmsv_est},
the entanglement estimate (middle curve) according to
Fig.~\protect\ref{estimate}, and the distance measure
(lower curve) according to Fig.~\protect\ref{distrel}
for the mean photon number \mbox{$\bar{n}$ $\!=$ $1$}
(\mbox{$|q|$ $\!\simeq$ $\!0.7071$}).
}
\end{figure}

In Fig.~\ref{abstandsvergleich} the entanglement degradation
as calculated in Section \ref{sec:distance} is compared with the
estimate obtained in Section \ref{sec:single} and the bound
obtained in Section \ref{sec:rains}.
The figure reveals
that the distance of the output state to the separable Gaussian
states (lower curve) is much smaller than it might be expected
from the bound on the entanglement (upper curve)
calculated according to Eq.~(\ref{2.30}) together with
Eqs.~(\ref{2.28}) and (\ref{3.20}), as well as
the estimate (middle curve) derived by extracting a 
single pure state according to Eq.~(\ref{3.15}).
Note that the entanglement of the single pure state (\ref{3.13})
comes closest to the distance of the actual state to the separable
Gaussian states, whereas
the convex sum (\ref{3.20}) of density operators in Schmidt
decomposition can give much higher values.
Both methods, however, overestimate the entanglement.
Since with increasing mean photon number
the convex sum contains more and more terms, the bound gets worse
[and substantially slower on the computer, whereas computation of the
distance measure (\ref{4.4}) does not depend on it].

Thus, in our view, the distance to the separable Gaussian states
should be the measure of choice for determining the entanglement
degradation of entangled Gaussian states. Nevertheless, it should be
pointed out that the distance to separable Gaussian states has been
considered and not the distance to all separable states. 
We have no proof yet, that there does not exist an inseparable
non-Gaussian state which is closer than the closest Gaussian state.

%%%%%%%%%%%%%%%%%%%%%%%%%%%%%%%%%%%%%%%%%%%%%%%%%%%%%%%%%%%%%%%%%%%%%%
\section{Conclusions}
\label{conclusions}

In the present article the interferometric generation and the
transmission of entangled light have been studied, with
special emphasis on Gaussian states. The optical devices such as
beam splitters and fibers are regarded as being dispersing
and absorbing dielectric four-port devices as typically used
in practice. In particular, their action on light is described in
terms of the experimentally measurable transmission, reflection,
and absorption coefficients.

An entangled two-mode state can be generated by mixing
single-mode non-classical light at a beam splitter. Depending on
the phases of the impinging light beams and the beam-splitter
transformation, the amount of entanglement contained in the
outgoing light can be controlled. For squeezed vacuum input states
and appropriately chosen phases, maximal entanglement is obtained
for lossless, symmetrical beam splitters.
In realistic experiments, however, losses such as material
absorption prevents one from realizing that value.

When entangled light is transmitted through optical devices, losses
give always rise to entanglement degradation. In particular, after
propagation of the two modes of a two-mode squeezed vacuum
through fibers the available entanglement can be drastically reduced.
Unfortunately, quantifying entanglement of mixed states
in an infinite-dimensional Hilbert spaces has been close to
impossible. Therefore, estimates and upper bounds for the entanglement
content have been developed.

The analytical estimate employed in this article is based on
extraction of a single pure state from the output
Gaussian state, using its reduced von
Neumann entropy as an estimate for the entanglement.
However, this method is neither unique, since there are many
different ways of extracting pure states, nor is it an upper bound,
since nothing is said about the residual entanglement contained in the
state which is left over. In principle, more and more pure
states could be extracted until the residual state becomes separable.

Instead, an upper bound can be calculated  by decomposing the
output Gaussian state in a convex sum of Schmidt states as proposed
in \cite{Hiroshima00}. The disadvantage of this method is that the
bound gets worse for increasing (statistical) mixing. In 
particular, it may give hints for large entanglement even if the
quantum state under consideration is almost separable.

In order to overcome the disadvantage,
the distance of the output Gaussian state to the set
of separable Gaussian states measured
by the relative entropy is considered. It has the advantage
that separable states obviously correspond to zero distance.
Although one has yet no proof that there does not exist a
non-Gaussian separable state which is closer to the Gaussian
state under consideration than the closest separable Gaussian
state, one has good reason to think that it is even an
entanglement measure. In any case, it is a much better bound
than the one obtained by convexity. In particular, it clearly
demonstrates the drastic decrease of entanglement of the output
state with increasing entanglement of the input state.
Moreover, one observes saturation of entanglement transfer;
that is, the amount of entanglement that can maximally
be contained in the output state is solely determined by
the transmission length and does not depend on the
amount of entanglement contained in the input state. 

%%%%%%%%%%%%%%%%%%%%%%%%%%%%%%%%%%%%%%%%%%%%%%%%%%%%%%%%%%%%%%%%%%%%%%
\acknowledgements
S.S. likes to thank V.I.~Man'ko for helpful discussions on
multivariable Hermite polynomials. The authors also acknowledge
discussions about Gaussian quantum states with E.~Schmidt.

%%%%%%%%%%%%%%%%%%%%%%%%%%%%%%%%%%%%%%%%%%%%%%%%%%%%%%%%%%%%%%%%%%%%%%

%%%%%%%%%%%%%%%%%%%%%%%%%%%%%%%%%%%%%%%%%%%%%%%%%%%%%%%%%%%%%%%%%%%%%%
\appendix
\section{Fock-state expansion of multimode squeezed vacuum states}
\label{app:a}

Let us consider an incoming field prepared
in the squeezed vacuum state (\ref{2.3}) and an
absorbing beam splitter in the ground state. 
The quantum-state transformation formula (\ref{2.20}) then leads to
\begin{eqnarray}
\label{A.2}
\hat{\varrho}_{\rm out}^{({\rm F})} = \sum\limits_{g_1,g_2=0}^\infty 
&& \langle g_1,g_2| \hat{S}_{a_1'}(\xi_1) \hat{S}_{a_2'}(\xi_2)
|0,0,0,0\rangle
\nonumber \\ &&
\langle 0,0,0,0| \hat{S}^\dagger_{a_2'}(\xi_2)
\hat{S}^\dagger_{a_1'}(\xi_1) |g_1,g_2\rangle
\end{eqnarray}
where the transformed operators $\hat{a}_i'$ are defined by
\begin{equation}
\label{A.3}
\hat{a}_i' = \sum\limits_{j=1}^4 \Lambda^\ast_{ji} \hat{a}_j
\end{equation}
according to the rules of quantum-state transformation.
Equivalently, the Fock-state expansion of the density matrix reads
\begin{eqnarray}
\label{A.4}
\lefteqn{
\langle m_1,m_2 | \hat{\varrho}_{\rm out}^{({\rm F})} | n_1,n_2
\rangle =}
\nonumber \\ &&
\langle m_1,m_2|\langle g_1,g_2| \hat{S}_{a_1'}(\xi_1)
\hat{S}_{a_2'}(\xi_2) |0,0,0,0\rangle
\nonumber \\ &&
\langle 0,0,0,0| \hat{S}^\dagger_{a_2'}(\xi_2)
\hat{S}^\dagger_{a_1'}(\xi_1) |g_1,g_2\rangle | n_1,n_2 \rangle \,.
\end{eqnarray}
Expanding the Fock states in terms of coherent states and
using the squeeze operator in the form given in the second
line in Eq.~(\ref{2.4}),
we obtain after performing all integrals
\begin{eqnarray}
\label{A.6}
\lefteqn{
\langle m_1,m_2 | \hat{\varrho}_{\rm out}^{({\rm F})} | n_1,n_2
\rangle =}
\nonumber \\[.5ex] &&
\sqrt{\frac{(1-|q_1|^2)(1-|q_2|^2)}{m_1!m_2!n_1!n_2!}}
(-1)^{m_1+m_2+n_1+n_2} 
\nonumber \\[.5ex] && \times
\sum\limits_{g_1,g_2=0}^\infty \frac{1}{g_1!g_2!}
\,H^{\bf M}_{m_1,m_2,g_1,g_2}({\bf 0})
H^{\ast{\bf M}}_{n_1,n_2,g_1,g_2}({\bf 0}),
\end{eqnarray}
where the Hermite polynomials of four variables are generated
by the symmetric matrix ${\bf M}$ with elements
\begin{equation}
\label{A.7}
M_{ij} = q_1\Lambda_{i1}\Lambda_{j1}+ q_2\Lambda_{i2}\Lambda_{j2}\,.
\end{equation}
Using the relation between Hermite polynomials of one variable and
those of several variables \cite{Erdelyi},
\begin{eqnarray}
\label{A.8}
\lefteqn{
\sum\limits_{m_1+\ldots +m_n=m} \frac{a_1^{m_1}}{m_1!} \cdots
\frac{a_n^{m_n}}{m_n!} H^{\bf M}_{m_1,\ldots,m_n}(x_1,\ldots.x_n)
}
\nonumber \\[.5ex] &&\hspace{15ex}
= \,\frac{[\textstyle\frac{1}{2}\phi({\bf a})]^{m/2}}{m!}
   \,H_m\!\left[
   \frac{\phi({\bf a},{\bf x})}{\sqrt{2\phi({\bf a})}} \right],
\end{eqnarray}
where
\begin{equation}
\label{A.8a}
\phi({\bf a},{\bf x}) = \sum_{i,j}a_iM_{ij}x_j
\end{equation}
and $\phi({\bf a})$ $\!\equiv$
$\!\phi({\bf a},{\bf a})$, we get for the multivariable Hermite
polynomial of zero argument
\begin{eqnarray}
\label{A.9}
\lefteqn{
H^{\bf M}_{m_1,m_2,g_1,g_2}({\bf 0}) = \frac{H_m(0)}{2^{m/2}}
\frac{m_1!m_2!g_1!g_2!}{m!}
}
\nonumber \\[.5ex] &&\hspace{8ex} \times
\sum_{\cal P} \underbrace{M_{i_1,j_1}M_{i_2,j_2}\cdots
M_{i_{m/2},j_{m/2}}}_{m/2\mbox{ terms}}
\qquad
\end{eqnarray}
($m$ $\!=$ $\!m_1+m_2+g_1+g_2$). Here, the ${\cal P}$-sum runs over all
$m!/(m_1!m_2!g_1!g_2!)$ possible combinations to distribute $m_1$
indices $1$, $m_2$ indices $2$, $g_1$ indices $3$, and $g_2$ indices
$4$ among the indices $i_1$, $j_1$, $\ldots$, $i_{m/2}$, $j_{m/2}$.

In particular, if we restrict ourselves to two dimensions,
Eq.~(\ref{A.9}) simplifies to
\begin{eqnarray}
\label{A.10}
\lefteqn{
H^{\bf M}_{m_1,m_2}({\bf 0}) =
}
\nonumber \\[.5ex] &&\hspace{1ex} 
\frac{H_m(0)}{2^{m/2}}
M_{11}^{(m_1-m_2-\nu)/4} M_{22}^{(m_2-m_1-\nu)/4} m_1!m_2!
\bigg(\frac{m}{2}\bigg)!
\nonumber \\[.5ex] &&\hspace{1ex} \times\!
\sum\limits_{n=\frac{1}{2}\nu}^{\left[\frac{\mu}{2}\right]} 
\frac{1}{n!(\mu-2n)!(n-\frac{1}{2}\nu)!}
M_{11}^n (2M_{12})^{\mu-2n} M_{22}^n
\end{eqnarray}
[$\mu$ $\!=$ $\max(m_1,m_2)$, $\nu$ $\!=$ $|m_1-m_2|$]. The sum can
also be calculated leading to Gegenbauer, Jacobi or associated
Legendre polynomials \cite{Dodonov94}.

Another way of writing is the one we used for the numerical
calculation of the density matrix elements. The method, however, is
only applicable in cases where the number of variables the Hermite
polynomial depends on is sufficiently small. The multivariable Hermite
polynomial (now in four variables) of zero argument can be written as
\begin{eqnarray}
\label{A.11}
\lefteqn{
(-1)^{m_1+m_2+g_1+g_2} H^{\bf M}_{m_1,m_2,g_1,g_2}({\bf 0}) =
}
\nonumber \\[.5ex] && \hspace{2ex}
\frac{\partial^{m_1+m_2+g_1+g_2}}{\partial \lambda_1^{m_1} \partial
\lambda_2^{m_2} \partial \lambda_3^{g_1} \partial \lambda_4^{g_2}}
\,\exp\!\left[-{\textstyle\frac{1}{2}}\mbb{\lambda}^{\rm T}
{\bf M}\mbb{\lambda}\right]_{\mbb{\lambda}={\bf 0}},
\end{eqnarray}
with ${\bf M}$ being given by Eq.~(\ref{A.7}). Expanding the rhs of
Eq.~(\ref{A.11}), the only surviving term is the one proportional to
$(\mbb{\lambda}^{\rm T}{\bf M}\mbb{\lambda})^{(m_1+m_2+g_1+g_2)/2}$.
Multinomial expansion of this term then leads
to Eq.~(\ref{2.23}).

%%%%%%%%%%%%%%%%%%%%%%%%%%%%%%%%%%%%%%%%%%%%%%%%%%%%%%%%%%%%%%%%%%%%%%
\section{Multimode Gaussian density operators and Wigner functions}
\label{app:b}

Given the Wigner function of an $N$-mode Gaussian
state of the form of
\begin{equation}
\label{B.1}
W_N(\mbb{\zeta})
= \frac{1}{(2\pi)^N \sqrt{\det {\bf V}}} \, \exp\!\left( 
- {\textstyle\frac{1}{2}} \mbb{\zeta}^{\rm T} {\bf V}^{-1}
\mbb{\zeta} \right),
\end{equation}
where \mbox{$\mbb{\zeta}$ $\!=$ $\!(x_1,p_1,\ldots,x_N,p_N)$}
is the $2N$-dimensional ``vector'' of the 
quadrature components of the $N$ (complex) variables
$\hat{a}_i$ and ${\bf V}$ is the $2N\times 2N$ variance matrix
of the quadrature components.
The characteristic function defined by
the Fourier transform reads
\begin{equation}
\label{B.2}
\chi_N(\mbb{\eta})
= \exp\!\left( -{\textstyle\frac{1}{2}} \mbb{\eta}^{\rm T}
{\bf V} \mbb{\eta} \right) .
\end{equation}
Alternatively, the quantum state can can be given
by the density operator
\begin{equation}
\label{B.3}
\hat{\varrho}= \frac{\exp
\!\left[ -\frac{1}{2} \left( \hat{a}^\dagger\, \hat{a} \right)
{\bf M} \displaystyle{\hat{a} \choose \hat{a}^\dagger} \right]}
{{\rm Tr}\left\{ \exp \left[ -\left(\hat{a}^\dagger \,\hat{a}\right)
{\bf M} \displaystyle{\hat{a} \choose \hat{a}^\dagger}
\right] \right\}} 
\end{equation}
(where $\hat{a}$ is actually an $N$-dimensional ``vector'' with
``components'' $\hat{a}_i$).

In order to relate the matrix ${\bf M}$ to the
matrix ${\bf V}$, we introduce a unitary transformation
\begin{equation}
{\hat{a} \choose \hat{a}^\dagger}'
   = \hat{U} {\hat{a} \choose \hat{a}^\dagger} \hat{U}^{-1}
   = {\bf U} {\hat{a} \choose \hat{a}^\dagger},
\end{equation}
where the matrix ${\bf U}$ is chosen such that it diagonalizes
${\bf M}$, hence \mbox{${\bf U}^+{\bf M}{\bf U}$ $\!=$
$\!\mbb{\Theta}$} (with $\mbb{\Theta}$ being diagonal).
Note, that ${\bf U}$ satisfies the generalized
unitary relation
\begin{equation}
\label{B.4}
{\bf UJU}^+ = {\bf J} \qquad \mbox{with} \quad
{\bf J}=\mbox{diag}\,({\bf I}_N,-{\bf I}_N) .
\end{equation}
Then, the characteristic function of the density operator
(\ref{B.2}) is
\begin{eqnarray}
\label{B.5}
\lefteqn{
\chi_N(\mbb{\lambda},\mbb{\lambda}^\ast) =
{\rm Tr}\left[\hat{\varrho} \hat{D}(\mbb{\lambda}) \right]
}
\nonumber \\[.5ex]&&
= {\rm Tr}\left\{ \hat{\varrho} \exp
   \left[ \left(\hat{a}^\dagger \,\hat{a}\right)
   {\mbb{\lambda} \choose -\mbb{\lambda}^\ast} \right] \right\}
\nonumber \\[.5ex]&& 
= {\rm Tr}\left\{ \hat{U}\hat{\varrho}\hat{U}^{-1}\hat{U}
   \exp\!\left[ \left(\hat{a}^\dagger\, \hat{a}\right)
{\mbb{\lambda} \choose -\mbb{\lambda}^\ast} \right]
\hat{U}^{-1} \right\}
\nonumber \\[.5ex]&& 
   = \exp\!\left[ -{\textstyle\frac{1}{2}}
   {\mbb{\lambda} \choose -\mbb{\lambda}^\ast }^+
   {\bf U} \left( {\textstyle\frac{1}{2}}
   \coth {\textstyle\frac{1}{2}} \mbb{\Theta} \right) 
   {\bf U}^+ {\mbb{\lambda} \choose -\mbb{\lambda}^\ast }
   \right]
\nonumber \\[.5ex]&& 
= \exp\! \left[ -{\textstyle\frac{1}{2}}
{\mbb{\lambda} \choose \mbb{\lambda}^\ast }^+ {\bf D}
{\mbb{\lambda} \choose \mbb{\lambda}^\ast } \right]
\end{eqnarray}
with an obvious definition of the matrix ${\bf D}$, thus establishing a 
relation between the matrix ${\bf M}$ in the exponential of the
density operator and the matrix ${\bf D}$ in the exponential of the
characteristic function. {F}rom the third to the fourth line in
Eq.~(\ref{B.5}) we have used the expression for the characteristic
function of a thermal state \cite{Gardiner}. In due course, the
normalization of the density operator is obtained as
\begin{equation}
\label{B.6}
{\cal N} = \prod\limits_{i=1}^N 2\sinh \frac{\Theta_i}{2} \,.
\end{equation}

The above description shows a way to compute the entropy of a Gaussian 
quantum state $\hat{\varrho}$ as well as the relative entropy between
two Gaussian quantum states $\hat{\varrho}$ and $\hat{\sigma}$ as
\begin{eqnarray}
\label{B.7}
{\rm Tr}\left( \hat{\varrho} \ln \hat{\varrho}\right)
   &=& \sum\limits_{i=1}^N
   \ln\!\left( 2\sinh \frac{\Theta_i}{2} \right)
   -\frac{1}{2} \,
   {\rm Tr}\left( {\bf M}_\varrho {\bf D}_\varrho\right),
\\
{\rm Tr}\left( \hat{\varrho} \ln \hat{\sigma}\right)
   &=& \sum\limits_{i=1}^N
   \ln\! \left( 2\sinh \frac{\vartheta_i}{2} \right)
   -\frac{1}{2}\, {\rm Tr}\left( {\bf M}_\sigma {\bf D}_\varrho\right) ,
\end{eqnarray}
where the $\Theta_i$ and $\vartheta_i$ are respectively the
eigenvalues of ${\bf M}_\varrho$ and ${\bf M}_\sigma$.

%%%%%%%%%%%%%%%%%%%%%%%%%%%%%%%%%%%%%%%%%%%%%%%%%%%%%%%%%%%%%%%%%%%%%%
\newpage

\newpage


\begin{thebibliography}{99}

\bibitem{Korolkova00}
N.~Korolkova and G.~Leuchs, {\it Multimode Quantum Correlations},
to appear in {\it Coherence and Statistics of Photons and Atoms}, ed. by
J.~Pe\v{r}ina, to be published by J.~Wiley and Sons, Inc.

\bibitem{Duan00a}
L.-M.~Duan, G.~Giedke, J.I.~Cirac, and P.~Zoller,
Phys. Rev. Lett. {\bf 84}, 2722 (2000).

\bibitem{Duan00b}
L.-M.~Duan, G.~Giedke, J.I.~Cirac, and P.~Zoller, Phys. Rev. A
{\bf 62}, 032304 (2000).

\bibitem{Parker00}
S.~Parker, S.~Bose, and M.B.~Plenio, Phys. Rev. A {\bf 61}, 032305
(2000). 

\bibitem{Braunstein98}
S.L.~Braunstein and H.J.~Kimble, Phys. Rev. Lett. {\bf 80}, 869 (1998).

\bibitem{Vedral98}
V.~Vedral and M.B.~Plenio, Phys. Rev. A {\bf 57}, 1619 (1998).

\bibitem{Hiroshima00}
T.~Hiroshima, Phys. Rev. A {\bf 63}, 022305 (2001).

\bibitem{Yurke86}
B.~Yurke, S.L.~McCall, and J.R.~Klauder, Phys. Rev. A {\bf 33}, 4033
(1986). 

\bibitem{Prasad87}
S.~Prasad, M.O.~Scully, and W.~Martienssen, Opt. Commun. {\bf 62}, 139
(1987).

\bibitem{Ou87}
Z.Y.~Ou, C.K.~Hong, and L.~Mandel, Opt. Commun. {\bf 63}, 118 (1987).

\bibitem{Fearn87}
H.~Fearn and R.~Loudon, Opt. Commun. {\bf 64}, 485 (1987).

\bibitem{Campos89}
M.A.~Campos, B.E.A.~Saleh, and M.C.~Teich, Phys. Rev. A {\bf 40}, 1371 
(1989).

\bibitem{Leonhardt93}
U.~Leonhardt, Phys. Rev. A {\bf 48}, 3265 (1993).

\bibitem{Wodkiewicz85}
K.~W\'{o}dkiewicz and J.H.~Eberly, J. Opt. Soc. Am. B {\bf 2}, 458 (1985).

\bibitem{Ma90}
X.~Ma and W.~Rhodes, Phys. Rev. A {\bf 41}, 4625 (1990).

\bibitem{Knoll99}
L.~Kn\"oll, S.~Scheel, E.~Schmidt, D.-G.~Welsch, and A.V.~Chizhov,
Phys. Rev. A {\bf 60}, 4716 (1999).

\bibitem{Scheel00c}
S.~Scheel, T.~Opatrn\'{y}, and D.-G.~Welsch, Paper presented at
the International Conference on Quantum Optics 2000, Raubichi,
Belarus, May 28-31, 2000, arXiv: {\it quant-ph/0006026}.

\bibitem{Buch}
L.~Kn\"oll, S.~Scheel, and D.-G.~Welsch,
{\em QED in dispersing and absorbing dielectric media}, to appear in
{\it Coherence and Statistics of Photons and Atoms}, ed. by
J.~Pe\v{r}ina, to be published by J.~Wiley and Sons, Inc.,
arXiv: {\it quant-ph/0006121}.

\bibitem{Rains99}
E.M.~Rains, Phys. Rev. A {\bf 60}, 179 (1999).

\bibitem{Wu00}
S.~Wu and Y.~Zhang, Phys. Rev. A {\bf 63}, 012308 (2001).

\bibitem{Gruner96}
T.~Gruner and D.-G.~Welsch, Phys. Rev. A {\bf 54}, 1661 (1996).

\bibitem{Scheel00a}
S.~Scheel, L.~Kn\"oll, T.~Opatrn\'{y}, and D.-G.~Welsch,
Phys. Rev. A {\bf 62}, 043803 (2000).

\bibitem{Chizhov01}
A.V.~Chizhov, E.~Schmidt, L.~Kn\"oll, and D.-G.~Welsch,
J. Opt. B {\bf 3}, 1 (2001).

\bibitem{Simon00}
R.~Simon, Phys. Rev. Lett. {\bf 84}, 2726 (2000).

\bibitem{Lee00}
J.~Lee, M.S.~Kim, and H.~Jeong, Phys. Rev. A {\bf 62}, 032305 (2000).

\bibitem{Erdelyi}
A.~Erdelyi, W.~Magnus, F.~Oberhettinger, and F.G.~Tricomi,
{\it Higher Transcendental Functions}, Vol.~2, (McGraw--Hill, New York, 
1953).

\bibitem{Dodonov94}
V.V.~Dodonov and V.I.~Man'ko, J. Math. Phys. {\bf 35}, 4277 (1994).

\bibitem{Gardiner}
C.W.~Gardiner, {\it Quantum Noise} (Springer, Berlin, 1991).

\end{thebibliography}
\end{document}